\documentclass{ws-ijmpc}
\usepackage[super,compress]{cite}
\usepackage{ulem}
\usepackage{hyperref}
\usepackage{lipsum,verbatim}
\usepackage{graphicx}
\usepackage{multirow,rotate,subfigure}
\usepackage{amscd,amsmath,amssymb}
\usepackage{booktabs}

\usepackage{enumitem}

\usepackage{adjustbox}

\usepackage{mathrsfs}
\usepackage{mathtools}

\usepackage{amscd,amssymb}
\usepackage{mathtools}

\usepackage{csquotes}

\usepackage{color, colortbl}
\definecolor{red}{rgb}{1,0,0}
\definecolor{green}{rgb}{0,1,0}
\definecolor{blue}{rgb}{0,0,1}
\definecolor{Gray}{gray}{0.9} 
\definecolor{LightCyan}{rgb}{0.88,1,1}
\definecolor{ciano}{rgb}{0,1,1}
\definecolor{magenta}{rgb}{1,0,1}
\definecolor{amarelo}{rgb}{1,1,0}
\definecolor{bananayellow}{rgb}{1.0, 0.88, 0.21}

\usepackage{color, colortbl}
\definecolor{babypink}{rgb}{0.96, 0.76, 0.76}
\definecolor{amarelo}{rgb}{0.91, 0.84, 0.42}  
\definecolor{aqua}{rgb}{0.0, 1.0, 1.0}  
\definecolor{iw}{rgb}{0.7,0.93,0.36}  
\definecolor{lav}{rgb}{0.96,0.73,1.0}  
\definecolor{awesome}{rgb}{1.0, 0.13, 0.32}    
\definecolor{cadetgrey}{rgb}{0.57, 0.64, 0.69} 
\definecolor{x11gray}{rgb}{0.75, 0.75, 0.75}   
\definecolor{bananayellow}{rgb}{1.0, 0.88, 0.21}
\definecolor{mistyrose}{rgb}{0.9,0.82,0.70}
\newcommand{\pk}{\cellcolor{babypink}}
\newcommand{\iw}{\cellcolor{iw}}
\newcommand{\iy}{\cellcolor{lav}}

\newcommand{\bl}{\cellcolor{aqua}}

\newcommand{\xg}{\cellcolor{x11gray}}
\newcommand{\mr}{\cellcolor{mistyrose}}
\newcommand{\by}{\cellcolor{bananayellow}}

\begin{document}

\markboth{Jason A.C.~Gallas}
{Preperiodicity and systematic extraction of periodic orbits
              of the quadratic map}

\catchline{}{}{}{}{}

\title{Preperiodicity and systematic extraction of periodic orbits
              of the quadratic map}

\vspace{-0.25truecm}

\author{Jason A.C.~Gallas} 

\address{
Instituto de Altos Estudos da Para\'\i ba,
  Rua Silvino Lopes 419-2502,\\   
  58039-190 Jo\~ao Pessoa, Brazil,\\
Complexity Sciences Center, 9225 Collins Ave.~Suite 1208, 
Surfside FL 33154, USA,\\
  Max-Planck-Institut f\"ur Physik komplexer Systeme,
  01187 Dresden, Germany\\
     jason.gallas@gmail.com}

\maketitle

\begin{history}
\received{21 August 2020}
\accepted{27 August 2020}
\centerline{https://doi.org/10.1142/S0129183120501740}
\end{history}


\begin{abstract}
Iteration of the quadratic map produces sequences of  polynomials whose
degrees {\sl explode} as the orbital period grows more and more.
The polynomial mixing all 335 period-12 orbits has degree $4020$, while
for the $52,377$ period-20 orbits the degree rises already to $1,047,540$.
Here, we show how to use preperiodic points 
to systematically extract exact equations of motion, one by one,
with no need for iteration.
Exact orbital equations provide valuable insight about the arithmetic
structure and nesting properties of towers of algebraic numbers which
define orbital points and bifurcation cascades of the map.
\keywords{Arithmetic dynamics; Preperiodic points; Quadratic map;
          Exact orbital equations.}
\end{abstract}

\ccode{PACS Nos.:
      02.10.De, 
      03.65.Fd, 
      02.70.Wz} 


\section{Introduction}

A recent paper in this Journal\cite{deco}
described an iterative approach to detect
orbit within orbit stratification in the so-called $a=2$
{\sl partition generating limit}\cite{bk1,bk2,bk3}
of the quadratic or, equivalently, logistic map
\begin{equation}
  x_{t+1} \equiv f(x_t) =  a - x_t^2,  \qquad\quad t=0, 1, 2, \dots.
       \label{qm}
\end{equation}
Orbit within orbit
stratification means that periodic orbits are not always necessarily
independent of each other.
As discussed in Refs.~\cite{deco,jg19}\;, apart from the intrinsic interest
in detecting interdependent orbits,
stratification is potentially significant to, e.g., rearrange trajectories
sums in trace formulas underlying a semiclassical interpretation of
atomic physics spectra.

The existence of stratification was originally detected by extracting
orbits from polynomials of very high degrees  obtained by iterating
Eq.~(\ref{qm}).
However, map iteration generates sequences of mammoth polynomials whose
degrees explode as the orbital period grows more and more.
To bypass this difficulty it was conjectured that exploration of preperiodic
points could provide a viable alternative to iteration.

Here, the aim is to complement the work of Ref.~\cite{deco}
by introducing and explicitly implementing an alternative algorithm
based on preperiodic points.
As it is known\cite{prep}, for $a=2$ preperiodic points
are easy to obtain. They are roots of an infinite family of
polynomials $Q_\ell(x)$ generated by a recursive relation,
given below in Eq.~(\ref{reco}).
Starting from preperiodic points one quickly lands in periodic cycles.
So far, preperiodic points were used\cite{prep} to extract specific orbital
equations embedded in polynomial clusters with degree exceeding
one billion and, consequently, totally out of reach by ordinary
brute-force polynomial factorization.
Here, in contrast, the aim is to investigate whether or not the roots
of $Q_\ell(x)$ are able to generate systematically, one by one,
{\sl all orbits} of the map.
We find that they are, as described in what follows.

\section{Motivation  to perform exact analytical work}

For any given period $k$, equations of motion are defined by $k$-degree
polynomials which have either {\sl integer\/}  or {\sl algebraic\/}
numbers as coefficients.
Thus, the algebraic nature of equations of motion depends of the
algebraic character of the polynomial coefficients representing the orbit.
By way of illustration, consider the following triplet of exact period-five
orbits\cite{deco}:
\begin{eqnarray}
o_{5,1}(x) &=& {x}^{5}-{x}^{4}-4\,{x}^{3}+3\,{x}^{2}+3\,x-1,\label{o51}\\
o_{5,2}(x) &=& {x}^{5}+\tfrac{1}{2}(1+\sqrt {33}) {x}^{4}-{x}^{3}
                 -\tfrac{3}{2}(3+\sqrt {33}) {x}^{2}-(6+\sqrt {33}) x-1,\label{o52}\\
o_{5,3}(x) &=& {x}^{5}+\tfrac{1}{2}(1-\sqrt {33}) {x}^{4}-{x}^{3}
             -\tfrac{3}{2}(3-\sqrt {33}) {x}^{2}-(6-\sqrt {33}) x-1.\label{o53}
\end{eqnarray}
Technically, $o_{5,2}(x)$ and $o_{5,3}(x)$ are {\sl conjugated} over the
quadratic number field $\mathbb Q(\sqrt{33})$.
When multiplied together, they produce an orbital {\it cluster\/}
$c_{5,1}(x)$, namely
\begin{eqnarray}
  c_{5,1}(x) &=& o_{5,2}(x)\cdot o_{5,3}(x),\cr
            &=& {x}^{10}+{x}^{9}-10\,{x}^{8}-10\,{x}^{7}+34\,{x}^{6}
                +34\,{x}^{5}-43\,{x}^{4}-43\,{x}^{3}+12\,{x}^{2}+12\,x+1.
\end{eqnarray}
Manifestly, this orbital aggregate  has integer coefficients and, therefore,
is arithmetically simpler than the pair of orbits that it contains.
While to multiply known conjugate orbits is an easy task, the inverse problem,
to disentangle conjugated orbits from a given cluster, is a
quite hard problem, particularly for aggregates of orbits with odd or
high degrees.
Iteration produces a profusion of orbital aggregates that need
to be disentangled.

Notice that since it is  not possible to represent $\sqrt{33}$ numerically
without truncation, the coefficients in $o_{5,2}(x)$ and $o_{5,3}(x)$ can only
be represented numerically as approximations.
This means that numerical (inexact) work precludes recognizing orbital
conjugations such as the symmetric decompositions clearly visible between
$o_{5,2}(x)$ and $o_{5,3}(x)$. 
Nevertheless, as shown below, knowing that orbital conjugation exists
one can suitably sift and multiply inexact orbits, searching for 
expressions which turn out to have nearly integer coefficients.
This is the key idea  to be explored in the remainder of the paper.

As mentioned,
the big challenge is to derive exact expressions for orbits  when
the orbital period $k$ grows without bound.
A complication hampering such derivation is that current computer algebra
systems are essentially adapted to deal with  procedures developed for
integers, not for generic algebraic numbers of arbitrary  degrees,
a considerably harder problem.
This motivates pursuing the alternative procedure discussed here,
which profits from preperiodic points of the $Q_\ell(x)$ family. 

\section{Generation of the $Q_\ell(x)$ polynomials}

The $Q_\ell(x)$ polynomials are obtained as irreducible factors of an
auxiliary family of polynomials, $T_\ell(x)$, which are generated recursively.
Starting from two initial {\sl seed  functions}, $T_0(x)$ and $T_1(x)$,
subsequent $T_\ell(x)$ are obtained from the recurrence\cite{equi}
\begin{equation}
  T_\ell(x) = x T_{\ell-1}(x) - T_{\ell-2}(x), \qquad 
             \ell=  2, 3, 4, \dots.   \label{reco}
\end{equation}
For our present purpose, we fix $T_0(x)=2$ and $T_1(x)=x$.
Instead of the recurrence, a direct way to obtain $T_{\ell}(x)$ 
is from  Pincherle's relation\cite{pin}
\begin{equation}
 T_{\ell}(x) =  \left(\frac{x-\sqrt{x^2-4}}{2}\right)^{\ell}
             + \left(\frac{x+\sqrt{x^2-4}}{2}\right)^{\ell},
      \qquad \ell=0, 1, 2, \dots.
\end{equation}

{\small
\begin{table}[!tbh]
\tbl{Characteristics of the orbits produced by the roots of the first
  120 polynomials $Q_\ell(x)$.
  The label $\ell$ refers to $Q_\ell(x)$ while
  $\partial_\ell$ gives the degree of the $Q_\ell(x)$
  which has the shortest preperiodic transient.
  Here,  $l$ is the length of the preperiodic transient leading
  to the orbit, and $k$ is the orbital period.
}
{\begin{tabular}{@{}|c|c|c|c|c|@{\hskip 0.3in}| c|c|c|c|c|@{\hskip 0.3in}| c|c|c|c|c|@{}} 
\hline
\hline
$i$ & $\ell$ & $\partial_\ell$ &$l$ & $k$ & $i$ & $\ell$ & $\partial\ell$
&$l$ & $k$ & $i$ & $\ell$ & $\partial_\ell$ &$l$ & $k$\\
\hline
    1&   2&   2&   3&   1&    41& 120&  64&   5&   4&    81& 258& 168&   3&   7\\
    2&   3&   2&   2&   1&    42& 136& 128&   5&   4&    82& 254& 252&   3&   7\\
    3&   4&   4&   4&   1&    43& 240& 128&   6&   4&    83& 344& 336&   5&   7\\
    4&   6&   4&   3&   1&    44& 272& 256&   6&   4&    84&  51&  32&   2&   8\\
    5&   8&   8&   5&   1&    45&  11&  10&   2&   5&    85&  85&  64&   2&   8\\
    6&  12&   8&   4&   1&    46&  22&  20&   3&   5&    86& 102&  64&   3&   8\\
    7&  16&  16&   6&   1&    47&  33&  20&   2&   5&    87& 170& 128&   3&   8\\
    8&  24&  16&   5&   1&    48&  31&  30&   2&   5&    88& 255& 128&   2&   8\\
    9&  32&  32&   7&   1&    49&  44&  40&   4&   5&    89& 257& 256&   2&   8\\
   10&  48&  32&   6&   1&    50&  66&  40&   3&   5&    90& 340& 256&   4&   8\\
\hline
   11&  64&  64&   8&   1&    51&  62&  60&   3&   5&    91& 408& 256&   5&   8\\
   12&  96&  64&   7&   1&    52&  88&  80&   5&   5&    92&  19&  18&   2&   9\\
   13& 128& 128&   9&   1&    53& 132&  80&   4&   5&    93&  27&  18&   2&   9\\
   14& 192& 128&   8&   1&    54& 124& 120&   4&   5&    94&  38&  36&   3&   9\\
   15& 256& 256&  10&   1&    55& 176& 160&   6&   5&    95&  54&  36&   3&   9\\
   16&   5&   4&   2&   2&    56& 264& 160&   5&   5&    96&  57&  36&   2&   9\\
   17&  10&   8&   3&   2&    57& 248& 240&   5&   5&    97&  73&  72&   2&   9\\
   18&  20&  16&   4&   2&    58& 352& 320&   7&   5&    98&  76&  72&   4&   9\\
   19&  40&  32&   5&   2&    59&  13&  12&   2&   6&    99& 108&  72&   4&   9\\
   20&  80&  64&   6&   2&    60&  21&  12&   2&   6&   100& 114&  72&   3&   9\\
\hline
   21& 160& 128&   7&   2&    61&  26&  24&   3&   6&   101& 171& 108&   2&   9\\
   22& 320& 256&   8&   2&    62&  42&  24&   3&   6&   102& 146& 144&   3&   9\\
   23&   7&   6&   2&   3&    63&  63&  36&   2&   6&   103& 152& 144&   5&   9\\
   24&   9&   6&   2&   3&    64&  52&  48&   4&   6&   104& 216& 144&   5&   9\\
   25&  14&  12&   3&   3&    65&  65&  48&   2&   6&   105& 228& 144&   4&   9\\
   26&  18&  12&   3&   3&    66&  84&  48&   4&   6&   106& 342& 216&   3&   9\\
   27&  28&  24&   4&   3&    67& 126&  72&   3&   6&   107& 304& 288&   6&   9\\
   28&  36&  24&   4&   3&    68& 104&  96&   5&   6&   108& 432& 288&   6&   9\\
   29&  56&  48&   5&   3&    69& 130&  96&   3&   6&   109& 437& 396&  86&   9\\
   30&  72&  48&   5&   3&    70& 168&  96&   5&   6&   110&  25&  20&   2&  10\\
\hline
   31& 112&  96&   6&   3&    71& 252& 144&   4&   6&   111&  41&  40&   2&  10\\
   32& 144&  96&   6&   3&    72& 208& 192&   6&   6&   112&  50&  40&   3&  10\\
   33& 224& 192&   7&   3&    73& 260& 192&   4&   6&   113&  93&  60&   2&  10\\
   34& 288& 192&   7&   3&    74& 336& 192&   6&   6&   114&  82&  80&   3&  10\\
   35&  15&   8&   2&   4&    75& 416& 384&   7&   6&   115& 100&  80&   4&  10\\
   36&  17&  16&   2&   4&    76&  43&  42&   2&   7&   116& 164& 160&   4&  10\\
   37&  30&  16&   3&   4&    77&  86&  84&   3&   7&   117& 200& 160&   5&  10\\
   38&  34&  32&   3&   4&    78& 129&  84&   2&   7&   118& 205& 160&   2&  10\\
   39&  60&  32&   4&   4&    79& 127& 126&   2&   7&   119& 341& 300&   2&  10\\
   40&  68&  64&   4&   4&    80& 172& 168&   4&   7&   120& 328& 320&   5&  10\\
\hline
\end{tabular}}\label{tab:tab02}
\end{table}
}

For $\ell=1$, $T_1(x)=Q_1(x)=x$.
For $\ell>1$, the polynomials $T_\ell(x)$ are always given by products
of  {\sl cyclotomic-like} irreducible factors $Q_\ell(x)$, except for
 $\ell=2^n$, $n=1,2,3,\dots$ when $T_\ell(x)=Q_\ell(x)$.
Every new $T_\ell(x)$ generated by  Eq.~(\ref{reco})
contributes a new irreducible factor $Q_\ell(x)$, new in the sense of
not appearing for any index $\ell'$ smaller than $\ell$.
Thus, the  first  few are $T_1(x)=Q_1(x)=x$, $T_2(x)=Q_2(x)=x^2-2$, and
\[  T_3(x)=Q_1(x)Q_3(x),\quad T_4(x)=Q_4(x), \quad T_5(x)=Q_1(x)Q_5(x),
            \quad T_6(x)=Q_2(x)Q_6(x),\]
where
\[  Q_3(x)=x^2-3,\quad Q_4(x)=x^4-4x^2+2, \quad Q_5(x)=x^4-5x^2+5,
          \quad Q_6(x)=x^4-4x^2+1.  \]
The first twenty  $T_\ell(x)$ and  $Q_\ell(x)$ are listed in Table~1 of an
open access paper\cite{prep}.
The key observation is that the {\it irreducible} $Q_\ell(x)$ are the
building blocks of the {\it reducible} auxiliary $T_\ell(x)$.
Using the roots of  $Q_\ell(x)$ as starting conditions
to iterate the quadratic map, Eq.~(\ref{qm}), one finds that after a
preperiodic start, i.e.~a certain number of non-repeating iterates, the
iteration lands on a cycle of $k$ distinct points that repeats forever.

\section{The selective extraction of periodic orbits}

Using Eq.~(\ref{reco}) we generated the first 400 polynomials $T_\ell(x)$ and
$Q_\ell(x)$.
Irreducible $T_\ell(x)$ imply $Q_\ell(x)=T_\ell(x)$. Otherwise,
$Q_\ell(x)$ is the factor of highest degree in $T_\ell(x)$.
Therefore, while the degree of $T_\ell(x)$ grows steadily with $\ell$,
the degree of $Q_\ell(x)$ fluctuates, i.e.~emerges not in a regular order.
After generating the 400 $Q_\ell(x)$ polynomials, we investigated
in which periodic orbit their roots land.
Clearly, there are several roots to choose as initial conditions of the
iterative process.
Thus, in addition to roots leading to genuine period-$k$ orbits one may
also find roots leading to orbits of smaller periods, divisors of $k$.

Table \ref{tab:tab02} illustrates data for 120 of the 400 polynomials,
ordered according the period $k$.
This table  reveals interesting systematic patterns and trends  which
will be considered in more detail now.

\subsection{Preperiodic generation of period four orbits and aggregates}

As it is known, there are three possible period-four orbits for the
quadratic map, namely
\begin{eqnarray}
o_{4,1}(x) &=& {x}^{4}+{x}^{3}-4\,{x}^{2}-4\,x+1,\\
o_{4,2}(x) &=&{x}^{4} -\tfrac{1}{2}(1-\sqrt {17}) {x}^{3}
    -\tfrac{1}{2}(3+\sqrt {17}) {x}^{2}  - (2+\sqrt {17}) x-1,\label{o42}\cr
    &\simeq& {x}^{4}+ 1.561553\,{x}^{3}- 3.561553\,{x}^{2}- 6.123106\,x- 1,\\
o_{4,3}(x)  &=& {x}^{4} -\tfrac{1}{2}(1+\sqrt {17}) {x}^{3}
   - \tfrac{1}{2}(3-\sqrt {17}) {x}^{2}  - (2-\sqrt {17}) x-1,\label{o43}\cr
    &\simeq& {x}^{4}- 2.561553\,{x}^{3}+ 0.561553\,{x}^{2}+ 2.123106\,x- 1.
\end{eqnarray}
Here, we have also indicated approximate  ``projections'' onto the real
axis for two orbits. This was done to emphasize that,
since no exact representation onto the real axis is possible for $\sqrt{17}$,
independently of the number of digits used,
such projections will be necessarily just {\it approximations\/}
of the exact equations of motions, obliterating completely the 
conjugation symmetry between  $o_{4,2}(x)$ and $o_{4,3}(x)$.
Nevertheless, knowing that $o_{4,2}(x)$ and $o_{4,3}(x)$ are
conjugated naturally lead us to multiply them together to obtain
an approximate equation for a cluster and,
after rounding off coefficients, the corresponding exact expression.
Explicitly, using the approximated orbits, we find:
\begin{eqnarray}
c_{4,1}(x)&=& o_{4,2}(x) \cdot o_{4,3}(x), \cr
   &\simeq&{x}^{8}- {x}^{7}- 7.000001\,{x}^{6}+ 6.000004\,{x}^{5}+ 15{x}^{4}
               - 10.00001\,{x}^{3}- 10{x}^{2}+ 4x+ 1,\cr
   &=& {x}^{8}-{x}^{7}-7{x}^{6}+6{x}^{5}+15{x}^{4}
               -10{x}^{3}-10{x}^{2}+4x+1, \qquad \Delta=17^7. \label{eq11}
\end{eqnarray}
The discriminant $\Delta$ corroborates that Eq.~(\ref{eq11}) decomposes
over $\mathbb Q(\sqrt{17})$, as it should.
The algebraic character of the coefficients of the three period-four orbits
forms two groups, according to the algebraic nature, integer or quadratic,
of the roots of
\begin{equation}
  \mathbb S_4(\sigma) = (\sigma+1)(\sigma^2-\sigma-4), \label{s4}
\end{equation}
where $\sigma$ is the sum of the orbital points.
When the root $\sigma=-1$ is substituted into the period-four carrier
$\psi_4(x)$, defined by  Eq.~(1) in Ref.~\cite{deco}, namely
\begin{eqnarray}
 \psi_4(x) &=&{x}^{4}-\sigma{x}^{3} + \tfrac{1}{2}({\sigma}^{2}+\sigma-8){x}^{2}
  -\tfrac{1}{6} ({\sigma}^{3}+3\,{\sigma}^{2}-20\,\sigma+2) x\cr
  &&\quad\quad\ +\tfrac{1}{24} (\sigma-3)( {\sigma}^{3}
       +9\,{\sigma}^{2}-2\,\sigma-16), \label{o4}
\end{eqnarray}
one obtains $o_{4,1}(x)$.
Substituting $(1-\sqrt{17})/2$ and $(1+\sqrt{17})/2$ into Eq.~(\ref{o4}),
roots of the quadratic factor in Eq.~(\ref{s4}),
we get $o_{4,2}(x)$ and $o_{4,3}(x)$, the pair of orbits conjugated over
$\mathbb Q(\sqrt{17})$.
For details, see Refs.~\cite{deco,jg19}\;.

\begin{table}[!tbh]
\centering
\tbl{The selective factorization
  of period-four orbits. See text.\label{tab:table04aa}}
{\begin{tabular}{@{}| l | ccc ccc ccc c| @{}} 
\hline
Polynomial & $Q_{15}$ &\bl $Q_{17}$  & $Q_{30}$  &\bl $Q_{34}$  & $Q_{60}$ &\bl $Q_{68}$
    & $Q_{120}$ &\bl $Q_{136}$  & $Q_{240}$  &\bl $Q_{272}$ \\
\hline
Degree & 8 &\bl 16& 16& \bl32&  32&\bl 64& 64&
         \bl   128& 128&\bl 256\\
\hline
Orbits & $o_{4,1}$&\bl $o_{4,2}$& $o_{4,1}$&\bl $o_{4,2}$& $o_{4,1}$&\bl $o_{4,2}$
         & $o_{4,1}$&\bl $o_{4,2}$& $o_{4,1}$&\bl $o_{4,2}$\\
     &     &\bl $o_{4,3}$&   &\bl $o_{4,3}$& &
          \bl$o_{4,3}$&  & \bl$o_{4,3}$&   &\bl $o_{4,3}$\\
\hline
\end{tabular}}\label{tab:tab04}
\end{table}

{\small
\begin{table}[!tbh]
  \tbl{The six period-five orbits in existence for
    the quadratic map, characterized by one orbital point,
    and by the sum $\sigma_{5,\ell}$ of its points. The remaining
    points follow by iterating  $x_{t+1}=2-x_t^2$.}
{\begin{tabular}{@{}|c||c|c|@{}} 
\hline
Orbit &  $x_1$ &  $\sigma_{5,\ell}$ \\
\hline
$o_{5,1}$ & -1.6825070656623623377 & 1\\
\hline
$o_{5,2}$ & -1.9638573945254134021 & -3.3722813232690143300\\
$o_{5,3}$ & -1.1601138191423963584 &  2.3722813232690143300\\
\hline
$o_{5,4}$ & -1.9590598825049889879 & -3.0838723594356076658\\
$o_{5,5}$ & -1.6415268824145526527 &  0.7868018150723329561\\
$o_{5,6}$ & -1.0579280206539249147 &  3.2970705443632747098\\
\hline
\end{tabular}}\label{tab:tab0a}
\end{table}}

{\small
\begin{table}[!tbh]
\centering
\tbl{The selective factorization of period-five orbits.
    \label{tab:table05aa}}
{\begin{tabular}{@{} | ccc ccc ccc ccc c | @{}} 
\hline
$Q_{11}$&$Q_{22}$&\bl $Q_{33}$&\by $Q_{31}$&$Q_{44}$&\bl $Q_{66}$%
&\by $Q_{62}$&$Q_{88}$&\bl $Q_{132}$&\by $Q_{124}$%
&$Q_{176}$&\bl $Q_{264}$&\by $Q_{248}$\\
\hline
10 & 20&\bl 20& \by30&  40&\bl 40&\by 60&
            80& \bl80&\by 120&   160& \bl160&\by 240\\
\hline
$o_{5,1}$& $o_{5,1}$&\bl $o_{5,2}$&\by $o_{5,4}$& $o_{5,1}$&\bl $o_{5,2}$
         & \by$o_{5,4}$& $o_{5,1}$&\bl $o_{5,2}$&\by $o_{5,4}$
         & $o_{5,1}$&\bl $o_{5,2}$&\by$o_{5,4}$ \\    
        &  &\bl $o_{5,3}$&\by $o_{5,5}$&  &\bl $o_{5,3}$&
\by$o_{5,5}$&  & \bl$o_{5,3}$&\by $o_{5,5}$&     &\bl $o_{5,3}$ &\by$o_{5,5}$\\    
        &  &    &\by $o_{5,6}$&    &   &
\by$o_{5,6}$&  &    &\by $o_{5,6}$&     & &\by$o_{5,6}$ \\    
\hline
\end{tabular}}\label{tab:tab0b}
\end{table}}

Thus, we see that even modest numerical knowledge of the orbital coefficients
can disclose the exact expression of the cluster equation.
This procedure is a significant asset when searching for
exact expressions for clusters aggregating orbits of high periods:
it allows one to profit from approximate numerical information to
correctly extract exact equations.

Table \ref{tab:tab04} shows the $Q_\ell(x)$ polynomials that generate
period-four orbits for $\ell\leq400$.
The topmost line identifies the polynomials while the second line
refers to their degrees.
Under them are indicated the orbits $o_{4,j}$ where the zeros of
$Q_\ell(x)$ land.
As illustrated by the highlighting, it is not difficult to recognize that
there are two nested sequences of period-four generating $Q_\ell(x)$
polynomials, namely
\[ Q_{15\times 2^n}(x) \qquad\hbox{and}\qquad Q_{17\times 2^n}(x),
     \qquad n=0,1,2,3,\dots. \]

Table \ref{tab:tab04} reveals the following  remarkable facts:
\begin{enumerate}[nolistsep]
\item[i)] All three period-four orbits in existence are generated by the
   zeros of $Q_\ell(x)$, in a cyclic manner;
\item[ii)] While orbits generated by $Q_{15\times 2^n}(x)$ consistently land
    on $o_{4,1}(x)$, the orbit with integer coefficients,
    orbits generated by $Q_{17\times 2^n}(x)$ land either on $o_{4,2}(x)$
    or $o_{4,3}(x)$, orbits which have quadratic numbers as coefficients.
This means that orbits belonging to the same number field are
{\it segregated automatically}, filtered, by the zeros of $Q_\ell(x)$;
\item[iii)] For a given period $k$, the coefficient of $x^{k-1}$
  in the orbital equation
is $-\sigma$. This fact provides easy access to  the equation defining the
{\sl sum of the orbital points}\cite{deco} which, in the present example,
turns out to given by Eq.~(\ref{s4}).
\end{enumerate}

As discussed in the continuation, these three features are found to
be generic characteristics of other $Q_\ell(x)$ polynomials and periods.

{\small
\begin{table}[!tbh]
\centering
\tbl{Selective factorization of orbits
  of periods six, seven, and eight.}
{\begin{tabular}{@{}| ccc ccc ccc ccc c| @{}} 
\hline
$Q_{13}$&$\pk Q_{21}$& $Q_{26}$&\pk $Q_{42}$&\by $Q_{63}$& $Q_{52}$%
&\iy $Q_{65}$&\pk $Q_{84}$&\by $Q_{126}$& $Q_{104}$%
&\iy$Q_{130}$&\pk $Q_{168}$&\by $Q_{252}$\\
\hline
12 &\pk 12& 24&\pk 24 &\by 36&  48&\iy 48&\pk 48&
           \by 72& 96&\iy 96& \pk 96& \by144\\
\hline
$o_{6,1}$&\pk $o_{6,2}$& $o_{6,1}$&\pk $o_{6,2}$&\by $o_{6,3}$& $o_{6,1}$
         &\iy$o_{6,6}$&\pk $o_{6,2}$&\by $o_{6,3}$& $o_{6,1}$
&\iy $o_{6,6}$&\pk $o_{6,2}$&\by $o_{6,3}$ \\
        &  &  &  &\by  $o_{6,4}$  & &
\iy$o_{6,7}$&  & \by$o_{6,4}$&    &\iy $o_{6,7}$&  &\by $o_{6,4}$\\    
        &  &    &  &\by$o_{6,5}$   &   &
\iy$o_{6,8}$&  & $\by o_{6,5}$   & &\iy $o_{6,8}$     & &\by$o_{6,5}$ \\    
        &  &    &  &  &  &\iy $o_{6,9}$  &
        &  &    &\iy $o_{6,9}$  & & \\    
\hline
\hline
\by$Q_{43}$&$\by Q_{86}$ &\iw $Q_{129}$&\xg $Q_{127}$&\by $Q_{172}$%
&\iw $Q_{258}$&\xg $Q_{254}$ &\by $Q_{344}$& && &&\\
\hline
\by42 &\by 84&\iw 84&\xg 126 &\by 168&\iw  168&\xg 252 &\by 336& && &&\\
\hline
\by$o_{7,1}$&\by $o_{7,1}$&\iw $o_{7,4}$&\xg $o_{7,10}$&\by $o_{7,1}$&\iw $o_{7,4}$
 &\xg $o_{7,10}$ &\by $o_{7,1}$& && &&\\
\by$o_{7,2}$&\by $o_{7,2}$&\iw $o_{7,5}$&\xg $o_{7,11}$&\by $o_{7,2}$&\iw $o_{7,5}$
 &\xg $o_{7,11}$ &\by $o_{7,2}$& && &&\\
\by$o_{7,3}$&\by $o_{7,3}$&\iw $o_{7,6}$&\xg $o_{7,12}$&\by $o_{7,3}$&\iw $o_{7,6}$
 &\xg $o_{7,12}$ &\by $o_{7,3}$& && &&\\
        &&           \iw  $o_{7,7}$&\xg $o_{7,13}$&            &\iw $o_{7,7}$
&\xg $o_{7,13}$ && && &&\\
        &&            \iw $o_{7,8}$&\xg $o_{7,14}$&            &\iw $o_{7,8}$
&\xg $o_{7,14}$ && && &&\\
        &&            \iw $o_{7,9}$&\xg $o_{7,15}$&            &\iw $o_{7,9}$
&\xg $o_{7,15}$ && && &&\\
        &&                     &\xg $o_{7,16}$&            & 
&\xg $o_{7,16}$ && && &&\\
        &&                     &\xg $o_{7,17}$&            & 
&\xg $o_{7,17}$ && && &&\\
        &&                     &\xg $o_{7,18}$&            & 
 &\xg $o_{7,18}$ && && &&\\
\hline
\hline
$Q_{51}$&$\iy Q_{85}$ & $Q_{102}$&\iy $Q_{170}$&\bl $Q_{255}$%
& \mr$Q_{257}$&\iy $Q_{340}$ & & && &&\\
\hline
 32 &\iy 64& 64&\iy 128 &\bl 128&\mr  256&\iy 256 & & && &&\\
\hline
 $o_{8,1}$&\iy $o_{8,3}$& $o_{8,1}$&\iy $o_{8,3}$&\bl $o_{8,7}$
&\mr $o_{8,15}$  &\iy $o_{8,3}$ & & && &&\\
 $o_{8,2}$&\iy $o_{8,4}$& $o_{8,2}$&\iy $o_{8,4}$&\bl $o_{8,8}$
     &\mr $o_{8,16}$  &\iy $o_{8,4}$ & & && &&\\
    &\iy $o_{8,5}$&   &\iy $o_{8,5}$&\bl $o_{8,9}$
     &\mr $o_{8,17}$  &\iy $o_{8,5}$ & & && &&\\
    &\iy $o_{8,6}$&   &\iy $o_{8,6}$&\bl $o_{8,10}$
     &\mr$o_{8,18}$  &\iy $o_{8,6}$ & & && &&\\
    & &   & &\bl $\vdots$
     &\mr$\vdots$  & & & && &&\\
    &   &   & &\bl $o_{8,14}$
     &\mr$o_{8,30}$  &  & & && &&\\
\hline
\end{tabular}}\label{tab:tab06b}
\end{table}}

\subsection{Preperiodic generation of period five orbits}

For period-five,  preperiodic points  orderly generate the six orbits 
recorded in Table \ref{tab:tab0a},  where they are 
characterized by one orbital point as well as
by the sum $\sigma_{5,\ell}$ of its five points.
The first few $Q_\ell(x)$ having roots which land
on period-five orbits are recorded in Table \ref{tab:tab0b}.
Similarly to Table \ref{tab:tab04},
the topmost line shows the relevant $Q_\ell(x)$, with their degrees 
on the second line.
For period-five, the highlighting shows the existence of
three distinct nested  sequences, namely:
\[ Q_{11\times 2^n}(x), \qquad Q_{33\times 2^n}(x), \qquad Q_{31\times 2^n}(x),
                     \qquad n=0,1,2,3,\dots. \]
Once again, the $Q_\ell(x)$ generate systematically all existing orbits in a
cyclic way, and segregate them automatically according to the algebraic nature
of the orbital coefficients.
As before, from the numerically obtained orbits we get the exact expression
defining  $\sigma_{5,\ell}$ for the six orbital points, namely
\begin{equation}
  \mathbb S_5(\sigma)=  (\sigma-1) ({\sigma}^{2}+\sigma-8)
               ({\sigma}^{3}-{\sigma}^{2}-10\sigma+8).
\end{equation}  
The three factors composing $\mathbb S_5(\sigma)$ correspond to the
three groups of orbits discriminated in Table \ref{tab:tab0b}.

From numerically approximate orbits we obtain exact expressions for the
pair of period-five clusters:
\begin{eqnarray}
c_{5,1}(x) &=& {x}^{10}+{x}^{9}-10\,{x}^{8}-10\,{x}^{7}+34\,{x}^{6}
              +34\,{x}^{5}-43\,{x}^{4}-43\,{x}^{3}+12\,{x}^{2}+12\,x+1,\\  
c_{5,2}(x) &=& {x}^{15}-{x}^{14}-14\,{x}^{13}+13\,{x}^{12}+78\,{x}^{11}
                 -66\,{x}^{10}-220\,{x}^{9}+165\,{x}^{8}+330\,{x}^{7}\cr
   &&\quad-210\,{x}^{6}-252\,{x}^{5}+126\,{x}^{4}+84\,{x}^{3} -28\,{x}^{2}-8\,x+1.
\end{eqnarray}
These aggregates factor into quintics over $\mathbb Q(\sqrt{33})$ and
$\mathbb Q\big(\sqrt[3]{-62+95\sqrt{-3}}\,\big)$, respectively,
thereby providing exact
explicit expressions for the remaining
five period-five orbits. Note that a {\it complex} number field is needed
to extract the three {\it real orbits} entangled in $c_{5,2}(x)$.

Once again, apart from $o_{5,1}(x)$, the real projections
of the five remaining orbits are necessarily approximated.
However, when multiplying together the distinct orbits arising from the
roots of a fixed $Q_\ell(x)$ polynomial we obtain a cluster whose 
coefficients turn out to be very close to integers.
Rounding them off yields the final exact expression with integer coefficients.
Even using modest numerical approximations of the orbital points one can
obtain exact cluster equations as may be validated by comparing its roots
with numerical values generated by iteration of the equations of
motion\cite{deco}.
We have encountered no case where the above procedure failed to produce
exact expressions for orbital clusters.

\subsection{Preperiodic generation of periods six, seven, and eight}

Altogether, there are nine orbits of period six, listed in Table~1 of
Ref.~\cite{deco}.
The orbits form four groups, corresponding to the four factors composing
$\mathbb S_6(\sigma)$, which defines the sum of their 
orbital points:
\begin{equation}
\mathbb S_6(\sigma) = (\sigma+1)(\sigma-1)( {\sigma}^{3}-21\,\sigma+28)
          ( {\sigma}^{4}+{\sigma}^{3}-24\,{\sigma}^{2}-4\,\sigma+16).
\end{equation}
Thus, there are two isolated orbits, denoted by $o_{6,1}(x)$ and $o_{6,2}(x)$,
which have integer coefficients, 
a group of three orbits with cubic coefficients, $o_{6,3}(x)$, $o_{6,4}(x)$,
and $o_{6,5}(x)$, and a group of four orbits with quartic coefficients,
$o_{6,6}(x)$, $o_{6,7}(x)$, $o_{6,8}(x)$, and $o_{6,9}(x)$.

The $Q_\ell(x)$ polynomials generating all period-six orbits are collected
in the upper portion of Table \ref{tab:tab06b}.
The center portion of the table collects the $Q_\ell(x)$ polynomials that
generate all period-seven orbits.
The three factors which compose $\mathbb S_7(\sigma)$, as well as the
procedure to obtain them, valid for any arbitrary period $k$,
are given explicitly in Ref.~\cite{deco}.
Finally, the lower portion of Table \ref{tab:tab06b} presents $Q_\ell(x)$
polynomials for all period-eight orbits.
Expressions for $\mathbb S_8(\sigma)$,  $\mathbb S_9(\sigma)$, and
$\mathbb S_{10}(\sigma)$ are  given in the Appendix.
The individual factors composing the several $\mathbb S_k(\sigma)$ fix the
algebraic character of the coefficients for every individual period-$k$ orbit.

\section{Conclusions and outlook}

This paper complements the iterative approach recently discussed in
this Journal\cite{deco}. Here, the aim was to obtain
an alternative method  to extract systematically exact 
expressions for orbital equations of arbitrary periods of the quadratic
map in the partition generating limit.
The alternative method consists of using preperiodic points
comfortably generated by an infinite family of {\sl monogenic}\cite{mono}
polynomials $Q_\ell(x)$ to selectively
extract equations of motion, one by one, with no need for iterating
polynomials.
The procedure is simple to implement and effective.

Both methods, polynomial iteration or preperiodic points,
are essentially limited by the capability of the  hardware and software
used to handle ever growing polynomials with huge numerical coefficients.
While this limitation impacts the maximum period accessible to algebraic
manipulations, continued advances in computer systems will certainly
continue to expand the range available to investigate algebraic
dynamics\cite{silver} exactly,
allowing one to advance into new research realms by applying any of the
two methodologies now available.

The quadratic map in Eq.~(\ref{qm}) offers a number of enticing problems
worth pursuing. For $a=0$, the orbits    reproduce several of
the familiar cyclotomic polynomials.
For $a=2$, the infinite set of periodic orbits embedded in the
fully developed chaos consists of a cyclotomic-like set of objects that
share many properties with the standard cyclotomic
polynomials\cite{deco,jg19}.
The dynamics for other values of $a$, when real and complex orbits coexist,
is totally open to investigation. Integer and rational values of $a$ are
first good candidates to learn how number towers unfold arithmetically.
In particular $a=1$, say, offers the
possibility of learning about the interplay of coexisting orbits
defined by towers of real and complex algebraic quantities, a new
and totally unexplored world.


{\small
\begin{table}[!bht]
  \tbl{The thirty period-eight orbits, characterized by one orbital point
    and the sum $\sigma_{8,j}$ of the eight orbital points.
    Complete orbits may be
    generated by iterating $x_{t+1}=2-x_t^2$. The values of $\sigma_{8,j}$
   are roots of Eq.~(\ref{s8}).}
{\begin{tabular}{@{}|c||c|c|@{}} 
\hline
Orbit &  $x_1$ &  $\sigma_{8,j}$ \\
\hline
$o_{8,1}$ & -1.984841019343871516522912 & -2.561552812808830274910705\\
$o_{8,2}$ & -1.632393824712443381743705 &  1.561552812808830274910705\\
\hline
$o_{8,3}$ & -1.994538346771576098442202 & -4.914223945039180928208247\\
$o_{8,4}$ & -1.951023935960873278145019 & -2.056133705669804629074231\\
$o_{8,5}$ & -1.573489876066966494406107 &  2.352671132230350653297542\\
$o_{8,6}$ & -1.738177892611056635019841 &  3.617686518478634903984936\\
\hline
$o_{8,7}$ & -1.999392903955743234132651 & -9.192789454613742069133553\\
$o_{8,8}$ & -1.970324466935013007529986 & -3.456817575142370539782937\\
$o_{8,9}$ & -1.898269888071802444878746 & 0.2899559343985875455768889\\
$o_{8,10}$ & -1.926986288411966237038570 & 0.5911182259052169094701404\\
$o_{8,11}$ & -1.687334295667532671943878 & 1.104146442912019886485395\\
$o_{8,12}$ & -1.784801166495895642936504 & 1.766177771271217083497342\\
$o_{8,13}$ & -1.224840406098499982124136 & 4.323105719133964018738106\\
$o_{8,14}$ & -0.9785858338678472056275101& 5.575102936135107165148616\\
\hline
 $o_{8, 15}$ & -1.999402315686187320311568 & -9.229152884143427069047449\\
 $o_{8, 16}$ & -1.994622984321411050382163 & -4.890484957292577579687509\\
 $o_{8, 17}$ & -1.970783422280053272462841 & -3.270791497771093661943518\\
 $o_{8, 18}$ & -1.985075746010103467347859 & -2.686687205494066985177061\\
 $o_{8, 19}$ & -1.867014790104666658874452 & -2.631302749249868342542760\\
 $o_{8, 20}$ & -1.951780177216468660609469 & -2.375150786131144323343685\\
 $o_{8, 21}$ & -1.899833775307932554727627 & 0.1093594181979987080920541\\
 $o_{8, 22}$ & -1.928111435796830339423029 & 0.7797117117494440466968254\\
 $o_{8, 23}$ & -1.637914468949626157934556 & 0.8210383845347267439370462\\
 $o_{8, 24}$ & -1.692053315444146323149561 & 1.834669950453959306149210\\
 $o_{8, 25}$ & -1.788077312909154517799568 & 1.949372205257687870932170\\
 $o_{8, 26}$ & -1.579860384419790809298828 & 2.955978611262725340347639\\
 $o_{8, 27}$ & -1.742147511485913313291649 & 3.175217183647612247623081\\
 $o_{8, 28}$ & -1.236026775830388280510551 & 3.430438713308632045309287\\
 $o_{8, 29}$ & -.9068916550266425682669635 & 4.627714403788193594030641\\
 $o_{8, 30}$ & -.9929341328818267363874723 & 6.400069497881198058624029\\
\hline
\end{tabular}}\label{tab:tab08}
\end{table}
}

\section*{Acknowledgments}
This work was started during a visit to the Max-Planck Institute for
the Physics of Complex Systems, Dresden, gratefully supported by an
Advanced Study Group on {\sl Forecasting with Lyapunov vectors}.
The author was partially supported by CNPq, Brazil, grant 304719/2015-3.

\appendix

\section{Expressions of $\mathbb S_k(\sigma)$ for periods  $k=8, 9$ and $10$}

Table \ref{tab:tab08} defines the thirty period-eight orbits $o_{8,i}(x)$
listed in Table \ref{tab:tab06b}.
The individual factors which fix the coefficients, and therefore
the algebraic nature for all period-eight orbits, are roots 
$\sigma_{8,j}$ of:
\begin{eqnarray}
 \mathbb S_8(\sigma) &=&  \big(\sigma^2+\sigma-4\big)
                \big(\sigma^4+\sigma^3-23\sigma^2-\sigma+86\big)\times\cr
 && \big({\sigma}^{8}-{\sigma}^{7}-75\,{\sigma}^{6}+261\,{\sigma}^{5}
    +474\,{\sigma}^{4}
   -2764\,{\sigma}^{3}+3560\,{\sigma}^{2}-1696\,\sigma+256\big)\times\cr
   && \big({\sigma}^{16}-{\sigma}^{15}-120\,{\sigma}^{14}+292\,{\sigma}^{13}+4390\,{\sigma}^{12}
   -13894\,{\sigma}^{11}-66604\,{\sigma}^{10}+257972\,{\sigma}^{9}\cr
 &&\qquad  +422785\,{\sigma}^{8}
   -2255633\,{\sigma}^{7}-434628\,{\sigma}^{6}+
   9169776\,{\sigma}^{5}-6074688\,{\sigma}^{4}\cr
   &&\qquad  -12553200\,{\sigma}^{3}+18123520\,{\sigma}^{2}
   -7237376\,\sigma+591872\big).  \label{s8}
\end{eqnarray}
Manifestly, there are four classes of orbital complexity, corresponding
to the factors of degrees $2, 4, 8, 16$.
See Ref.~\cite{deco} for details.

By constructing tables analogous to Table \ref{tab:tab08} we get expressions
for $\mathbb S_k(\sigma)$ which define the sum of orbital points for
periods $k=9$ and $10$, as well as the algebraic nature for all the
orbits of these periods.
From the $1+1+2+4+6+18+24=56$ period-nine orbits we find:
\begin{eqnarray}
  \mathbb S_9(\sigma) &=& \sigma \big(\sigma-1\big)\big({\sigma}^{2}+\sigma-14\big)
  \big({\sigma}^{4}-{\sigma}^{3}-27\,{\sigma}^{2}+41\sigma+2\big)\times\cr
 && \big({\sigma}^{6}-57\,{\sigma}^{4}+76\,{\sigma}^{3}+684\,{\sigma}^{2}-1824\sigma+1216\big)
  \times\big({\sigma}^{18}-171\,{\sigma}^{16}+342\,{\sigma}^{15}+9234\,{\sigma}^{14}\cr
  &&\qquad -25992\,{\sigma}^{13}
  -216030\,{\sigma}^{12}+707940\,{\sigma}^{11}+2274813\,{\sigma}^{10}-8209976\,{\sigma}^{9}
  -9918855\,{\sigma}^{8}\cr
  &&\qquad + 37877526\,{\sigma}^{7}+24342192\,{\sigma}^{6}-74783088\,{\sigma}^{5}
  -43931952\,{\sigma}^{4}+53908320\,{\sigma}^{3}\cr
  &&\qquad+40777344\,{\sigma}^{2}+2976768\sigma-622592\big)\times
  \big({\sigma}^{24}+{\sigma}^{23}-217\,{\sigma}^{22}+175\,{\sigma}^{21}
  +17702\,{\sigma}^{20}\cr
  &&\qquad -34450\,{\sigma}^{19}
  -713778\,{\sigma}^{18}+1978990\,{\sigma}^{17}+15416541\,{\sigma}^{16}
  -54321171\,{\sigma}^{15}\cr
  &&\qquad  -174437381\,{\sigma}^{14}+785332035\,{\sigma}^{13}+859244108\,{\sigma}^{12}
  -5882602892\,{\sigma}^{11}-234490112\,{\sigma}^{10}\cr
  &&\qquad  +20915101712\,{\sigma}^{9}   -7544198464\,{\sigma}^{8}
  -35643986496\,{\sigma}^{7}+15193333504\,{\sigma}^{6}\cr
  &&\qquad  +28736640000\,{\sigma}^{5}
  -9095376896\,{\sigma}^{4}-11068506112\,{\sigma}^{3}+1392246784\,{\sigma}^{2}\cr
  &&\qquad  +1681915904\sigma+134217728\big).
\end{eqnarray}
Apart from a pair of orbits with integer coefficients, there are five
classes of orbital complexity, corresponding to the remaining factors
in $\mathbb S_{9}(\sigma)$, and which define  orbital coefficients
of algebraic degrees $2, 4, 6, 18$, and $24$.

Analogously, from the $1+2+3+8+15+30+40 =99 $ period-ten orbits we get
the expressions that fix  coefficients and the  algebraic character
for all orbits:
\begin{eqnarray}
\mathbb S_{10}(\sigma) &=& \sigma \big({\sigma}^{2}-\sigma-10\big)
    \big({\sigma}^{3}+{\sigma}^{2}-10\sigma-8\big)
\big({\sigma}^{8}+{\sigma}^{7}-79\,{\sigma}^{6}+11\,{\sigma}^{5}+1766\,{\sigma}^{4}
             -1980\,{\sigma}^{3}\cr
&& -6120\,{\sigma}^{2}+6400\sigma+2560\big)\times
\big({\sigma}^{15}+{\sigma}^{14}-138\,{\sigma}^{13}+80\,{\sigma}^{12}+6278\,{\sigma}^{11}
 -13450\,{\sigma}^{10}\cr
&& -98056\,{\sigma}^{9}+360148\,{\sigma}^{8}+54921\,{\sigma}^{7}
 -1300271\,{\sigma}^{6}+477210\,{\sigma}^{5}+ 1783924\,{\sigma}^{4}\cr
&& -627480\,{\sigma}^{3} -972256\,{\sigma}^{2}+154496\sigma+141824\big)\times
 \big({\sigma}^{30}-{\sigma}^{29}-308\,{\sigma}^{28}+988\,{\sigma}^{27}\cr
&&  +35612\,{\sigma}^{26}-165388\,{\sigma}^{25}
 -2057832\,{\sigma}^{24}+12178568\,{\sigma}^{23}+64943174\,{\sigma}^{22}\cr
&& -488429574\,{\sigma}^{21}-
 1123435104\,{\sigma}^{20}+11745093392\,{\sigma}^{19}+9331519964\,{\sigma}^{18}\cr
&& -180278015100\,{\sigma}^{17}
 +3569846216\,{\sigma}^{16}+1838668414168\,{\sigma}^{15}-869341730175\,{\sigma}^{14}\cr
&& -12716885593921\,{\sigma}^{13}+9033287197044\,{\sigma}^{12}
 +59693364821364\,{\sigma}^{11}\cr
&& -48003427786304\,{\sigma}^{10}-186012055610544\,{\sigma}^{9}
 +147967410696768\,{\sigma}^{8}\cr
&& + 364183488569536\,{\sigma}^{7}-259735306624768\,{\sigma}^{6}
 -399991760098304\,{\sigma}^{5}\cr
&& +232462373875712\,{\sigma}^{4}+194622659919872\,{\sigma}^{3}
 -82496776568832\,{\sigma}^{2}\cr
&& -29669038489600\sigma+10950019121152\big)\times
 \big({\sigma}^{40}-410\,{\sigma}^{38}+820\,{\sigma}^{37}+69905\,{\sigma}^{36}\cr
&& -249444\,{\sigma}^{35}
 -6473900\,{\sigma}^{34}+32183360\,{\sigma}^{33}+354950530\,{\sigma}^{32}
 -2315420880\,{\sigma}^{31}\cr
&&-11590177004\,{\sigma}^{30}+102791354200\,{\sigma}^{29}+201751419530\,{\sigma}^{28}
 -2950337399160\,{\sigma}^{27}\cr
&& -577704999440\,{\sigma}^{26}+56031016855856\,{\sigma}^{25}
 -53560368567875\,{\sigma}^{24}-711891460448400\,{\sigma}^{23}\cr
&& +1295736491580950\,{\sigma}^{22}
+6043257616401300\,{\sigma}^{21}-15792646015357819\,{\sigma}^{20}\cr
&&-33550951726098500\,{\sigma}^{19}+119952431016405420\,{\sigma}^{18}
+113943905945026160\,{\sigma}^{17}\cr
&&-599301088769483360\,{\sigma}^{16}
-185763739671009920\,{\sigma}^{15}+1988903009578851200\,{\sigma}^{14}\cr
&&-104975203722073600\,{\sigma}^{13}-4318548194918598400\,{\sigma}^{12}
+1018330875056000000\,{\sigma}^{11}\cr
&&+5904641008147348480\,{\sigma}^{10}
-1616363148079616000\,{\sigma}^{9}-4743519951800729600\,{\sigma}^{8}\cr
&&+842331694258585600\,{\sigma}^{7}+1957033680587980800\,{\sigma}^{6}
+15775574654976000\,{\sigma}^{5}\cr
&&-276814761033728000\,{\sigma}^{4} -19653221731532800\,{\sigma}^{3}
+11527531161190400\,{\sigma}^{2}\cr
&&+792421466112000\sigma-112699941847040\big).
\end{eqnarray}
The above expression shows that there is one period-ten orbit with integer
coefficients, and six classes of orbital complexity, each class characterized
by coefficients defined by algebraic numbers of
degrees $2,3,8,15,30$, and $40$.

Next, there are 186 orbits of period $k=11$, 335 of $k=12$,
630 of $k=13$, 1161 of $k=14$, 2182 of $k=15$, etc\cite{bg14},
involving polynomial clusters of {degrees} $2046, 4020, 8190, 16254$,
and $32730$, respectively.
The algebraic properties of the orbits with $k>10$ remain to be investigated.
As shown in Refs.~\cite{deco,jg19}\;,
together with $\psi_k(x)$, the polynomials $\mathbb S_k(\sigma)$ form
``doublets'' defining {\it orbital carriers},
which {$\sigma$-encode} simultaneously all existing period-$k$ orbits
and, therefore, contain maximum possible information regarding the
complete set of period-$k$ orbits, for any arbitrary period $k$.





\begin{thebibliography}{00}

\bibitem{deco} J.A.C.~Gallas,
  Orbital carriers and inheritance in discrete-time quadratic dynamics,
  Int. J. Mod. Phys. C {\bf 31}, 2050100 (2020).
  
\bibitem{bk1} J.~Argyris, G.~Faust, M.~Haase, and R.~Friedrich,
  An Exploration of Dynamical Systems and Chaos, Second Edition
  (Springer, Berlin, 2015).

\bibitem{bk2} M.~Cencini, F.~Cecconi, and A.~Vulpiani,
  Chaos - From Simple Models to Complex Systems
  (World Scientific, Singapore, 2010).

\bibitem{bk3} M.~Ausloos and M.~Dirickx (eds.), 
      The Logistic Map: Map and the Route to Chaos: 
              From the Beginning to Modern Applications,
 Proceedings of the "Verhulst 200 on Chaos", Brussels, Belgium
 (Springer, Heidelberg, 2005).

\bibitem{jg19} J.A.C.~Gallas,   Lasers, stability, and numbers,
  Physica Scripta {\bf 94}, 014003 (2019). 
  
\bibitem{prep} J.A.C.~Gallas,
  Method for extracting arbitrarily large orbital equations of the
  Pincherle map,
  Results in Physics {\bf 6}, 561-567 (2016).

\bibitem{equi} J.A.C.~Gallas,
  Equivalence among orbital equations of polynomial maps,
   Int. J. Mod. Phys. C {\bf 29}, 1850082 (2018).
  
\bibitem{pin} S.~Pincherle,
   L'iterazione completa di $x^2-2$,
   Realle Accad. dei Lincei, Rend. della Classe di Scienze Fisiche,
   Matematiche e Naturali (Roma), Series 5, {\bf 29}(1), 329-333 (1920).

\bibitem{mono} J.A.C.~Gallas,  
  Monogenic period equations are cyclotomic polynomials,
  Int. J. Mod. Phys. C {\bf 31}, 2050058 (2020).

\bibitem{silver} J.H.~Silverman,
  The Arithmetic of Dynamical Systems
  (Springer, New York, 2007).

\bibitem{bg14} O.J.~Brison and J.A.C.~Gallas,
  What is the effective impact of the explosive orbital growth
  in discrete-time one-dimensional polynomial dynamical systems?
  Physica A {\bf 410},  313-318  (2014).   
  
\end{thebibliography}
\end{document}